%% file: main.tex
\documentclass{article}
\usepackage{ijcai21}

\usepackage{times}
\usepackage{soul}
\usepackage{url}
\usepackage[hidelinks]{hyperref}
\usepackage[utf8]{inputenc}
\usepackage[small]{caption}
\usepackage{graphicx}
\usepackage{amsmath}
\usepackage{amsthm}
\usepackage{booktabs}
\usepackage{algorithm}
\usepackage{algorithmic}
\usepackage[T1]{fontenc}
\usepackage{bm}
\usepackage{subfigure}
\usepackage{wrapfig}
\usepackage{helvet} 
\usepackage{courier}  
\usepackage{graphicx} 
\usepackage{caption} 
\usepackage{amssymb}
\usepackage{latexsym}
\usepackage{bbding}
\usepackage{booktabs}
\usepackage{multirow}
\usepackage{color}
\usepackage{authblk}

\newtheorem{theorem}{\bf Theorem}
\newtheorem{lemma}{\bf Lemma}

\newtheorem{definition}{\bf Definition}

\title{Improved Matrix Gaussian Mechanism for Differential Privacy\footnote{Work in progress}
}

\author[$ \ddagger $]{Jungang~Yang}
\author[$ \ddagger $]{Liyao~Xiang \thanks{Liyao Xiang (xiangliyao08@sjtu.edu.cn) is the corresponding author with the John Hopcroft Center, Shanghai Jiao Tong University, China.}}
\author[$ \ddagger $]{Weiting~Li}
\author[$ \ddagger $]{Wei~Liu}
\author[$ \ddagger $]{Xinbing~Wang}
\affil[$ \ddagger $]{Shanghai Jiao Tong University}

\begin{document}

\maketitle

\begin{abstract}
The wide deployment of machine learning in recent years gives rise to a great demand for large-scale and high-dimensional data, for which the privacy raises serious concern. Differential privacy (DP) mechanisms are conventionally developed for scalar values, not for structural data like matrices. Our work proposes Improved Matrix Gaussian Mechanism (IMGM) for matrix-valued DP, based on the necessary and sufficient condition of $ (\varepsilon,\delta) $-differential privacy. IMGM only imposes constraints on the singular values of the covariance matrices of the noise, which leaves room for design. Among the legitimate noise distributions for matrix-valued DP, we find the optimal one turns out to be i.i.d. Gaussian noise, and the DP constraint becomes a noise lower bound on each element. We further derive a tight composition method for IMGM. Apart from the theoretical analysis, experiments on a variety of models and datasets also verify that IMGM yields much higher utility than the state-of-the-art mechanisms at the same privacy guarantee.

\end{abstract}

\input{intro}

\input{prelim}

\input{formulation}

\input{method}

\input{comparison}
\input{experiment}

\input{related}
\input{conclusion}

\bibliographystyle{named}
\bibliography{main}
\newpage
\appendix

\input{appendix2}

\end{document}


\title{Supplemental Materials for A Distributional Robustness Certificate by Randomized Smoothing}
\author{}
\maketitle

\appendix
\input{appendix}
\input{appendix_B}
\bibliographystyle{named}
\bibliography{main}

%% file: intro.tex
\section{Introduction}

Proposed as a rigorous mathematical concept, differential privacy (DP) is renowned for preserving individual privacy over statistical results. Many data services today begin to adopt DP as a golden standard to protect personal information privacy in the data collection or processing. By introducing randomness through different means, most commonly, adding Laplace noise or Gaussian noise ~\cite{dwork2006calibrating,dwork2014algorithmic}, DP mechanisms return statistical results which are indistinguishable on the adjacent inputs. However, as the tension between utility and privacy widely exists, an overwhelming amount of noise is usually required in the high privacy regime, detrimental to results utility. Approaches have been taken to alleviate the tension, such as a refined privacy analysis on the higher moments of the privacy loss \cite{abadi2016deep}, or a sufficient and necessary condition for DP \cite{balle2018improving}. 

Matrix-valued data, or other high-dimensional data are frequently used today but there are few discussions on their DP mechanisms. An attempt is to apply scalar or vector based DP schemes to the matrix-valued data. However, matrices are different from scalars or vectors due to their intrinsic structural information. For example, the row of a matrix represents a data record whereas the column denotes an attribute. A DP mechanism may insert additional noise if it ignores the relation and simply treats each element individually.


An effort in applying DP to matrix-valued data is to insert matrix-variate Gaussian noise to achieve $(\varepsilon, \delta)$-differential privacy. It is found in MVG~\cite{chanyaswad2018mvg} that the DP condition only relates to the covariance matrices of the noise, and hence directional noise can be inserted to achieve better utility. However, similar to most existing Gaussian mechanisms, the matrix-variate approach is quite limited in practice, as the generated noise is too large to be of any use, especially in the high privacy regime ($\varepsilon \rightarrow 0$). The composition of MVG can be even more cumbersome. We observe that the fundamental cause is the application of a sufficient, rather than a necessary and sufficient condition of DP, leading to a smaller feasible range which limits the capability to find a good noise distribution.

We propose a practical DP mechanism for matrix-valued data, called Improved Matrix Gaussian mechanism (IMGM). By running a privacy loss analysis for matrix-valued data, we impose a necessary and sufficient condition of DP on the privacy loss variable. We first establish the condition on a particular input pair $X$ and $X^{\prime}$, then extend the condition to any pair of adjacent inputs. The necessary and sufficient condition of $(\epsilon, \delta)$-DP turns out to be an equivalent constraint on the smallest singular values of the noise covariance matrices, bounded by $\ell_{2}$-sensitivity of the query function. With an enlarged feasible range, we are able to greatly reduce the lower bound of the differentially-private noise. 

We claim that, for any additive Gaussian mechanism on matrix-valued data, IMGM provides the optimal noise distribution to achieve $(\epsilon, \delta)$-DP in terms of utility. It is optimal for two reasons: {\em first,} IMGM adopts a necessary and sufficient condition of DP to derive the noise bound, and thus seeks noise distributions in the largest possible feasible range; {\em second,} among the noise distributions meeting the DP condition, we find one that minimizes the magnitude of noise, incurring the least impact to the result utility. It is an interesting observation that the optimal Gaussian noise in the matrix setting is in fact i.i.d. Gaussian noise, having the same form with the scalar DP case. The conclusion is contradictory to the analysis of MVG that, directional noise does not bring additional utility benefit. Further, we also derive the composition theorem for IMGM by composing the privacy loss variables of independent IMGMs. The composition is tight since the necessary and sufficient condition of DP is applied.



We summarize highlights of our contributions as follows. 
\begin{enumerate}
	\item We propose Improved Matrix Gaussian Mechanism for matrix-valued queries, by deriving a noise bound over the necessary and sufficient condition of $ (\varepsilon,\delta) $-differential privacy.
	\item For all additive Gaussian mechanisms on matrix-valued data, we acquire the optimal noise distributions in terms of utility. A tight composition of IMGM is also proposed based on the necessary and sufficient condition.
	\item Experiments are conducted on a variety of tasks, models and datasets. The experimental results agree with our theory by showing IMGM has superior performance over prior mechnanisms in all settings.
\end{enumerate}

%% file: prelim.tex
\section{Previous Conclusions}
\label{sec:prelim}
We introduce existing definitions and theorems on the matrix-valued differential privacy mechanisms. 
\begin{definition}[$(\varepsilon,\delta)$-Differential Privacy]
	\label{def:dp}
	A randomized mechanism $K$ satisfies $ (\varepsilon,\delta) $-differential privacy if for any datasets $ {X} $ and $ {X}^{\prime} $ differing by at most one unit, and for any possible output $\mathcal{O}$,
	\begin{equation} \label{eq:dpeq}
	\Pr (K({X}) \in \mathcal{O}) \le e^{\varepsilon} \Pr (K({X}^{\prime}) \in \mathcal{O} ) +\delta.
	\end{equation}
\end{definition}
In the special case of $ \delta = 0 $ we call $ K ~ \varepsilon $-differentially private.

\begin{definition}[Matrix Gaussian distribution]\label{lemma:TVG}
	The probability density function for the $m \times n$ matrix-valued random variable $ {Z} $ which follows the matrix Gaussian distribution $ \mathcal{MN}_{m,n}(M,\Sigma_{1},\Sigma_{2}) $ is
	\begin{equation}\label{def:matrix gaussian}
	\Pr({{Z}}|{M},{\Sigma_1},{\Sigma_2}) = \frac{\exp\{-\frac{1}{2} \| U_1^{-1} ({Z}-M) U_2^{-\top}\|_F^2  \}}
	{(2\pi)^{(mn)/2}|{\Sigma_2}|^{n/2}|{\Sigma_1}|^{m/2}},
	\end{equation}
	where $ U_1\in \mathbb{R}^{m \times m}, U_2\in \mathbb{R}^{n \times n}  $ are invertible matrices and $ U_{k}U_{k}^{\top} = \Sigma_{k}, ~ k=1,2$. $ |\cdot| $ is the matrix determinant, ${M} \in \mathbb{R}^{m \times n}$ is mean, ${\Sigma}_1 \in \mathbb{R}^{m \times m}$ is the row-wise covariance and ${\Sigma_2} \in \mathbb{R}^{n \times n}$ is the column-wise covariance.
\end{definition}
Particularly, if $ N \in \mathbb{R}^{m \times n}$ is a Standard Normal Distribution (SND) random variable, every element of $ N $ follows the standard normal distribution $ \mathcal{N}(0, 1).$ Obviously, ${N} = U_1^{-1} ({Z}-M) U_2^{-\top},$ and ${N}$ is a special case that $ {N}\sim \mathcal{MN}_{m,n}(\bm{0}, \mathbf{E}_{1}, \mathbf{E}_{2})$ where $ \mathbf{E}_{1}, \mathbf{E}_{2} $ are the identity matrices of the same size of $ {\Sigma}_1  $ and $ \Sigma_{2} $ respectively.

Let $p_{K(X)}(Y)$ denote the probability density function of the random variable $Y=K(X),$ given mechanism $K$. The privacy loss function of $K$ on a pair of adjacent inputs $X \simeq X^{\prime}$ is defined as
\begin{equation}
\ell_{K, X, X^{\prime}}(Y)=\log \left(\frac{p_{K(X)}(Y)}{p_{K\left(X^{\prime}\right)}(Y)}\right).
\end{equation}
By the function $\ell_{K, X, X^{\prime}}$, the output random variable $Y=K(X)$ is transformed into the privacy loss random variable $L_{K, X, X^{\prime}}=\ell_{K, X, X^{\prime}}(Y)$. In particular, the Gaussian perturbation mechanism $K(X)=f(X)+Z$ with ${Z} \sim  \mathcal{MN}_{m,n}(\mathbf{0},\Sigma_{1},\Sigma_{2})$ has a privacy loss random variable which follows the Gaussian distribution:
\begin{lemma}\label{privacy loss}
	We define the privacy loss variable of mechanism $K(X)=f(X)+Z$ with ${Z} \sim  \mathcal{MN}_{m,n}(\mathbf{0},\Sigma_{1},\Sigma_{2})$ on the pair of inputs $X, X^{\prime}$ as $L_{K, X, X^{\prime}}.$ Then $L_{K, X, X^{\prime}} \sim \mathcal{N}(\eta, 2 \eta)$ with $\eta=\|U_1^{-1}\Delta U_2^{-\top}\|_F^{2} / 2,$ where $\Delta = f(X) - f(X^{\prime})$.
\end{lemma}
We provide its proof in Appendix~\ref{appendix: lemma 1}. A similar conclusion on the scalar value can be found in \cite{dwork2016concentrated}. 

\begin{lemma}\label{lemma: privacy loss iff DP}
	A mechanism $K: \mathbb{X} \rightarrow \mathbb{Y}$ is $(\varepsilon, \delta)$-differentially private  if and only if the following holds for every $X \simeq X^{\prime}:$
	\begin{equation}\label{eq: lemma privacy loss inequality}
	\Pr \left[L_{K, X, X^{\prime}} \geq \varepsilon\right]-e^{\varepsilon} \Pr \left[L_{K, X, X^{\prime}} \leq-\varepsilon\right] \leq \delta.
	\end{equation}
\end{lemma}
The conclusion can be found in \cite{balle2018improving} for vectorized values. Since its proof only concerns the probability distribution of the privacy loss variable, disregarding its shape, we can easily extend Lemma~\ref{lemma: privacy loss iff DP} to matrix-valued case. We will later take advantage of the property in our proofs.

%% file: formulation.tex
\section{Our Approach }
\label{sec:improvedbound}
As we found that, previous literature on matrix-valued differential privacy is intrinsically hard to apply in practice, as the inserted noise can be overwhelmingly large leading to significant utility decline. There are two reasons: the privacy loss variable is not well characterized in the matrix setting; the privacy analysis of the matrix-valued DP relies on the sufficient condition rather than the necessary and sufficient condition. In this section we introduce the necessary and sufficient condition for $(\varepsilon, \delta)$-DP on matrix-valued data, and present Improved Matrix Gaussian Mechanism (IMGM) based on it.
\subsection{Improved Matrix Gaussian Mechanism}
For a more fluent narrative of our mechanism, we first define $ l_{2} $-sensitivity on a pair of adjacent matrices (differing by a single record) as follows:
\begin{definition}[$ \ell_{2} $-sensitivity]
	\label{def:l2sensitivity}
	The $ \ell_{2} $-sensitivity of the query function $ f({X}) \in \mathbb{R}^{ m \times n } $ is defined as
	$$
	s_{2}(f) = \underset{d({X},{X}^{\prime}) = 1 }{\sup} \| f({X}) - f({X}^{\prime}) \|_{F},
	$$
	where $ \| \cdot \|_{F} $ is the Frobenius norm. 
\end{definition}

We apply the additive matrix Gaussian noise following $ \mathcal{MN}_{m,n}(\bm{0},\Sigma_{1}, \Sigma_{2}) $ distribution as follows.
\begin{definition} [Improved Matrix Gaussian Mechanism] \label{def:tvgm}
	For a given query function $ f({X}) \in \mathbb{R}^{m \times n}$ and a matrix variate Gaussian $ {Z} \sim  \mathcal{MN}_{m,n}(\mathbf{0},\Sigma_{1},\Sigma_{2}), $ the mechanism is defined as: 
	\begin{equation}
	\mathbf{IMGM}(f({X})) = f({X}) + {Z}.
	\end{equation}
\end{definition}


With zero-mean Gaussian noise added in IMGM, we present an equivalent condition of Eq.~\ref{eq: lemma privacy loss inequality} for a particular adjacent input pair $X, X^{\prime}$:
\begin{lemma}\label{lemma: IMGM}
	For any $\varepsilon \geq 0$, we define
	\begin{equation}\label{eq: Phi}
	g(x) = \Phi\left(\frac{x}{2}-\frac{\varepsilon }{x}\right)-e^{\varepsilon} \Phi\left(-\frac{x}{2 }-\frac{\varepsilon }{x }\right).
	\end{equation}
	For any given $\delta \in (0,1)$, the root of $ g(x) = \delta $ is $ B $.
	For adjacent input pair $ X $ and $ X^{\prime} $, let $ \Delta = f(X) - f(X^{\prime}) $, $ \Delta^{\prime} = U_1^{-1} \Delta U_2^{-\top}$ and $ U_k U_k^{\top} = \Sigma_{k},~k=1,2.$  We have the inequality~\eqref{eq: lemma privacy loss inequality} is equivalent to
	\begin{equation}\label{eq:delta'<B}
	\|\Delta^{\prime}\|_F \le B,
	\end{equation}
	for $ X $ and $ X^{\prime} $.
\end{lemma}

The full proof is in Appendix~\ref{appendix: lemma 3}. Since Eq.~\ref{eq: lemma privacy loss inequality} has been proved to be equivalent to Eq.~\ref{eq:dpeq}, we actually present the equivalence between Eq.~\ref{eq:dpeq} and Eq.~\ref{eq:delta'<B}. But note that the condition does not directly suggest differential privacy, as it associates with a particular adjacent input pair.

Given $B$, we further propose an equivalent condition for Lemma~\ref{lemma: IMGM} in terms of the singular values of $ U_1 $ and $ U_2 $:
\begin{theorem}\label{thm: IMGM}
	Let $f: \mathbb{X} \rightarrow \mathbb{R}^{m\times n}$ and $ \Delta = f(X) - f(X^{\prime}) $. With the same denotations of Lemma~\ref{lemma: IMGM}, we have $ \|\Delta^{\prime}\|_F \le B $ holds if
	\begin{equation}\label{eq:iff sigma}
	\sum_{i=1}^{r}\frac{\sigma_i^2(\Delta)}{\sigma_{m-i+1}^2(U_1)\sigma_{n-i+1}^2(U_2)} \le B^2
	\end{equation}
	for the particular $ X $ and $ X^{\prime} $, where $ r = \min\{m,n\} $. $\{ \sigma_i(A) \}$ is the non-increasingly ordered singular values of matrix $ A $. 
\end{theorem}
The full proof is in Appendix~\ref{appendix: thm 1}. Essentially, we prove by showing that $ \|\Delta^{\prime}\|_F$ is upper bounded by the left side of Eq.~\ref{eq:iff sigma}. Note that the sufficient condition is only related to the singular values of $ U_1 $ and $ U_2 $.

We observe that the necessary and sufficient condition (Eq.~\ref{eq: lemma privacy loss inequality}) needs to hold for every pair of $X \simeq X^{\prime}$ to meet the differential privacy guarantee. It essentially requires that for any $ \| \Delta \|_{F} \le s_{2}(f) $,
\begin{equation} \label{eq:maxdp}
\max_{\| \Delta \|_{F} \le s_{2}(f)} \|U_1^{-1}\Delta U_2^{-\top}\|_F \le B.
\end{equation}
Therefore we need to seek the upper bound of $ \|U_1^{-1}\Delta U_2^{-\top}\|_F $. Let
\begin{equation}
\Delta = W_{U_1} S_{\Delta} W_{U_2}^{\top},
\end{equation}
where $ S_{\Delta} = diag(\sigma_{1}(\Delta),\ldots, \sigma_{r}(\Delta)) $ is a diagonal matrix 
with the same shape of $ \Delta $. If we design $U_1 =  W_{U_1}S_{U_1} $ and $U_2 =  W_{U_2}S_{U_2}$, and substitute them into $ \|\Delta^{\prime}\|_F = $
\begin{equation}
\begin{split}
\|U_1^{-1}\Delta U_2^{-\top}\|_F & = \| S_{U_1}^{-1} W_{U_1}^{-1} W_{U_1} S_{\Delta} W_{U_2}^{\top} W_{U_2}^{-\top} S_{U_2}^{-1}\|_F \\
& = \|S_{U_1}^{-1}  S_{\Delta} S_{U_2}^{-1}\|_F \\
& = \sqrt{\sum_{i=1}^{r} \frac{\sigma_i^2(\Delta)}{\sigma_{m-i+1}^2(U_1)\sigma_{n-i+1}^2(U_2)}}.
\end{split}
\end{equation}
The third equality holds since $ W_{U_1}, W_{U_2} $ are orthogonal matrices. On the other hand, due to Eq.~\ref{eq:iff sigma}, it is proven that the upper bound of $ \|U_1^{-1}\Delta U_2^{-\top}\|_F $ can be achieved with careful design. 

Note that the left side of Eq.~\ref{eq:maxdp} should be without $\Delta$ when the optimal value is achieved. Actually, a set of noise would satisfy given the singular values of the covariance matrices. In particular, when $\sigma_{m-i+1}^2(U_1)\sigma_{n-i+1}^2(U_2) = C$ for $i \in \{1, \ldots, r \}$, the condition of Eq.~\ref{eq:maxdp} can be written more compactly, leading to the sufficient and necessary condition for differential privacy:

\begin{theorem}\label{thm: IMGM_new}
	Let $f: \mathbb{X} \rightarrow \mathbb{R}^{m\times n}$ and $ s_2(f) $ be its $ l_2 $-sensitivity. For any $\varepsilon>0$ and $\delta \in(0,1),$ the IMGM with $Z\sim  \mathcal{MN}_{m,n}(\mathbf{0},\Sigma_{1},\Sigma_{2})$ is $(\varepsilon, \delta)$-DP if and only if 
	\begin{equation}
	\begin{split}
	&\sigma_{m}(U_1) \sigma_{n}(U_2) \geq \frac{s_2(f)}{B}, \\
	\end{split}
	\end{equation}
	where $ U_k U_k^{\top} = \Sigma_{k},~k=1,2 .$ $\{ \sigma_i(A) \}_{i=1}^{n}$ is the non-increasingly ordered singular values of matrix $ A \in \mathbb{R}^{n \times n}$.  And $ B $ is the root defined in Lemma~\ref{lemma: IMGM}.
\end{theorem}
We present the full proof in Appendix~\ref{appendix: proof of thm IMGM new}. With Thm.~\ref{thm: IMGM_new}, IMGM could provide infinite noise distributions that meet the $ (\varepsilon,\delta) $-DP conditions, which admit many design choices considering the post-processing tasks.  Users can choose the noise distribution most suitable for their tasks. Since Thm.~\ref{thm: IMGM_new} only restricts the lower bound of the minimum singular value, we have the freedom to choose the values for other singular values of $ U_1,~U_2 $ beyond $\sigma_{m}(U_1), \sigma_{n}(U_2)$. From the aspect of minimizing the noise overhead, it would be a typical choice to choose other singular values the same as the minimum ones, {\em i.e.,} $ \sigma_{i}(U_1) = \sigma_{m}(U_1),~ \sigma_{j}(U_2)=\sigma_{n}(U_2) ,~\forall i\in [m], j\in [n]. $ Moreover, the directional matrices of $\Sigma_{1}$ (e.g. $ \Sigma_{1} = U_1 U_1^{\top} = W_{U_1} S_{U_1} S_{U_1}^{\top} W_{U_1}^{\top} $
, $ W_{U_1} $ is the directional matrix of $ \Sigma_{1} $) will not impact the guarantee of $ (\varepsilon,\delta) $-DP. In the case that we choose $ W_{U_k} = \mathbf{E}_{k}, k=1,2, $ we have
\begin{equation}\label{eq: choose U_1 U_2}
\begin{split}
&U_1 = \sigma_{m}(U_1) \mathbf{E}_1,\\
&U_2 = \sigma_{n}(U_2) \mathbf{E}_2.
\end{split}
\end{equation}
According to Def.~\ref{def:matrix gaussian}, $ Z = U_1 N U_2^{\top} $ with $ {N}\sim \mathcal{MN}_{m,n}(\bm{0}, \mathbf{E}_{1}, \mathbf{E}_{2})$. With Eq.~\ref{eq: choose U_1 U_2}, we could calculate $ Z = \frac{s_2(f)}{B} N$. Hence the optimal noise distribution for guaranteeing $(\epsilon, \delta)$-DP on matrix-valued data actually retreats to an i.i.d. Gaussian distribution. The detailed algorithm of IMGM is given in Alg.~\ref{alg: IMGM}.
\begin{algorithm}
	\renewcommand{\algorithmicrequire}{\textbf{Input:}}
	\renewcommand{\algorithmicensure}{\textbf{Output:}}
	\caption{Improved Matrix Gaussian Mechanism}
	\label{alg: IMGM}
	\begin{algorithmic}[1] 
		\REQUIRE (a) privacy parameters $\varepsilon,\delta$; (b) the query function and its sensitivity: $f(X) \in \mathbb{R}^{m \times n}, s_{2}(f) .$
		\STATE {Solve $ g(x) = \delta $ in Lemma~\ref{lemma: IMGM} to obtain the root $ B $ by the Alg.\ref{alg: privacy bound} in Appendix~\ref{appendix: alg to  Caculate $ B $}.} 
		\STATE { Draw a matrix-valued noise $$ Z = \frac{s_2(f)}{B}	\mathcal{MN}_{m,n}(\bm{0}, \mathbf{E}_{1}, \mathbf{E}_{2})$$ }
		\RETURN {$f(X)+Z $}
	\end{algorithmic} 
\end{algorithm}

\begin{theorem}
	Let $f: \mathbb{X} \rightarrow \mathbb{R}^{m\times n}$ and $ s_2(f) $ be its $ l_2 $-sensitivity. For any $\varepsilon>0$ and $\delta \in(0,1),$ Alg.~\ref{alg: IMGM} is $(\varepsilon, \delta)$-DP 
\end{theorem}

\cite{balle2018improving} implements a solver to seek the value of $B$ given a numerical oracle for computing $\Phi(t)$ based on the error function. We follow the same approach depending on binary search and the detail can be found in Appendix~\ref{appendix: alg to  Caculate $ B $}.

%% file: method.tex
\subsection{Composition}
Drawing on the privacy loss variable, we devise the composition theorem for IMGM. The idea is to compute the privacy loss variable for the composed mechanism, and apply the necessary and sufficient condition on that variable.

\begin{theorem}\label{thm: IMGM_Deep learning}
Suppose that a mechanism ${K}$ consists of a sequence of mechanisms ${K}_{1}, \ldots, {K}_{T}$ where ${K}_{t}: \prod_{j=1}^{t-1} \mathcal{R}_{j} \times \mathcal{D} \rightarrow \mathcal{R}_{t} .$ The composition result of the series of mechanisms $ K_{t}(X) = f(X) + Z $  with $Z\sim  \mathcal{MN}_{m,n}(\mathbf{0},\Sigma_{1},\Sigma_{2})$ is $(\varepsilon, \delta)-D P$ if and only if
	\begin{equation}\label{eq: Phi under moments account}
	\begin{split}
	&\sigma_{m}(U_1) \sigma_{n}(U_2) \geq \frac{s_2(f) \sqrt{T}}{B},~i \in [r] \\
	\end{split}
	\end{equation}
	where $ U_k U_k^{\top} = \Sigma_{k},~k=1,2 .$ $\{ \sigma_i(A) \}_{i=1}^{n}$ is the non-increasingly ordered singular values of matrix $ A \in \mathbb{R}^{n\times n}$. And $ B $ is the root defined in Lemma~\ref{lemma: IMGM}.
\end{theorem}

We present the full proof in Appendix~\ref{appendix: proof of deep learning with IMGM}.
Obviously, we can see that the per mechanism $\sigma_{m}(U_1) \sigma_{n}(U_2)$ is linear with $ \sqrt{T}.$ 

%% file: comparison.tex

\section{Comparison with MVG}
Desgined for matrix-valued queries, Matrix Variate Gaussian (MVG) mechanism $K(X)=f(X)+Z$ with ${Z} \sim  \mathcal{MN}_{m,n}(\mathbf{0},\Sigma_{1},\Sigma_{2})$ guarantees  $ (\varepsilon,\delta) $-differential privacy by imposing contraints on $ \Sigma_{1}$ and $ \Sigma_{2}$:
\begin{theorem}[MVG \cite{chanyaswad2018mvg}]
	\label{DP_MVG_Theo}
	Let 
	\[
	\begin{split}
	&{\sigma}( {\Sigma_1}^{-1}) = [ {\sigma}_{1}( {\Sigma_1}^{-1}),..., {\sigma}_{m}( {\Sigma_1}^{-1})]^{T},\\
	&{\sigma}( {\Sigma_2}^{-1}) = [ {\sigma}_{1}( {\Sigma_2}^{-1}),..., {\sigma}_{n}( {\Sigma_2}^{-1})]^{T},
	\end{split}
	\] 
	be the vectors of the non-increasingly ordered singular value of $ {\Sigma_1}^{-1}$ and $ {\Sigma_2}^{-1}$ respectively. The MVG mechanism guarantees $(\varepsilon,\delta)$-differential privacy if $ {\Sigma_1}$ and $ {\Sigma_2}$ satisfy the following condition:
	\begin{equation}
	\label{MVG_theorem_function}
	\| {\sigma}( {\Sigma_1}^{-1})\|_{2} \| {\sigma}( {\Sigma_2}^{-1})\|_{2} \le \frac{(-\beta_{0} + \sqrt{ \beta_{0}^{ 2} + 8\alpha_{0} \varepsilon })^{2}}{4 \alpha_{0}^{ 2}},
	\end{equation}
	where $\alpha_{0} = [H_{r}+H_{r,1/2}]\gamma^{2} + 2 H_{r} \gamma s_{2}(f) $, $ \beta_{0} = 2 (mn)^{1/4} H_{r} \zeta(\delta) s_{2}(f) $, $s_{2}(f)$ is the $\ell_{2}$-sensitivity of $f$, $\zeta(\delta)=2 \sqrt{-m n \ln \delta}-2 \ln \delta+m n$, $ \gamma =  {\sup}_{{X}} \| f({X}) \|_{F} $, $ r = \min\{m,n\} $ and $ H_r $ is generalized harmonic numbers of order $ r $.
\end{theorem}
Thm.~\ref{DP_MVG_Theo} mainly states that if the $ \ell_2 $-norm of the singular values of the inverse covariance matrices satisfy a given upper bound, $ (\varepsilon , \delta) $-DP can be satisfied. Therefore, MVG provides such an intuition that since only the sum of the values are constrained, a trade-off can be made on the singular values of different directions. Hence, MVG introduces the concept of Directional Noise: by designing the singular values in the noise distribution, different magnitudes of noise would be inserted to different directions, leading to varied utility results. An improved utility can be achieved by introducing less perturbation to the more important directions. And the importance of each direction is determined by domain knowledge or differentially-private SVD/PCA. While we appreciate the idea, we found it is still suboptimal.

\subsection{Why i.i.d Noise is Better}
Our approach has found that the DP condition essentially translates to constraints on the minimum value of the singular values of the noise covariance matrices. Hence, the i.i.d. noise would suffice for the optimal utility. There are two advantages: first, MVG establishes the constraints on $ \sigma({\Sigma_1}^{-1}) $ and $ \sigma({\Sigma_2}^{-1}) ,$ working in a smaller feasible region, while IMGM seeks the noise distribution in a larger feasible region, and thus finds better solutions. Second, MVG constrains the $\ell_{2}$-norm of the $\sigma$s and thus the directional noise may not impact its overall magnitude. IMGM improves such constraints to a $\ell_{\infty}$-like bound, of which any direction change would increase the noise overhead. An exception is that, one can still choose to tradeoff the row-wise noise and the column-wise noise by choosing different values of $\sigma_{m}(U_1)$ and $\sigma_{n}(U_2)$.

\subsection{Limitations in High Dimension and High Privacy Regime}
Recall the condition in Thm.~\ref{DP_MVG_Theo} and consider the case of increasing the values of $ m,n. $ Since $\zeta(\delta)=\Theta(mn)$ and $ \beta_{0} = \Theta(mn^{5/4}) ,$ we could rewrite the upper bound as:
\begin{equation}
	\resizebox{.91\linewidth}{!}{$
	\displaystyle
\frac{(-\beta_{0} + \sqrt{ \beta_{0}^{ 2} + 8\alpha_{0} \varepsilon })^{2}}{4 \alpha_{0}^{ 2}}  = \frac{16 \varepsilon^2}{(\beta_{0} + \sqrt{ \beta_{0}^{ 2} + 8\alpha_{0} \varepsilon })^{2}} = \Theta(mn^{-5/2}).
$}
\end{equation}
In the high dimension case that $ mn \rightarrow \infty $, we find that $ \| {\sigma}( {\Sigma_1}^{-1})\|_{2} \| {\sigma}( {\Sigma_2}^{-1})\|_{2} \rightarrow 0 $. And the variance of the Gaussian noise will increse to infinity, which will nullify the mechanism. Hence MVG cannot yield tight bounds for the regime that $ mn $ is large. But IMGM will provide an bound irrelevant to the matrix dimension, with a fixed $ s_2(f) $. This is because the equation $ g(x) = \delta $ does not change with dimensions, nor the root $ B, $ leading to constant variance for any dimension. Thus our approach is suitable to high-dimensional matrices.

When it comes to the high privacy regime, {\em i.e.,} $ \varepsilon \rightarrow 0 $, the limitations are similar to the high dimensional regime. We have that
\begin{equation}
	\frac{(-\beta_{0} + \sqrt{ \beta_{0}^{ 2} + 8\alpha_{0} \varepsilon })^{2}}{4 \alpha_{0}^{ 2}}  \rightarrow 0,
\end{equation}
when $ \varepsilon \rightarrow 0 $. Therefore, the variance provided by Thm.~\ref{DP_MVG_Theo} grows to infinity. This clearly falls outside the capabilities of MVG. But IMGM could still obtain a feasible variance from $ g(x) = \delta $ as  $ \varepsilon \rightarrow 0 $. The root $ B = 2 \Phi^{-1}(\delta/2+1/2) $ is also valid.

%% file: experiment.tex
\section{Experiments}

\begin{figure*}[t]\small
	\begin{minipage}[]{1\linewidth}
		\centering
		\makeatletter\def\@captype{table}\makeatother	
		\caption{Setup for Experiments. For gradients, only the one with the largest size is reported.}
		\label{tab:exp_setups}
		\setlength{\tabcolsep}{0.6mm}
\begin{tabular}{|c|c|c|c|c|c|c|}
	\hline
	Dataset        & MNIST               & CIFAR-10             & Cora                & Adult            & SVHN                 & IMDB               \\ \hline
	Model          & LeNet               & VGG-16               & GCN-Mixed           & MLP              & AlexNet              & BiLSTM             \\ \hline
	Training       & 55,000              & 50,000               & 1,140               & 32,561           & 73,257               & 25,000             \\ \hline
	Testing        & 5,000               & 10,000               & 1,000               & 16,281           & 26,032               & 25,000             \\ \hline
	Lot size       & 1,024               & 1,024                & 200                 & 512              & 1024                 & 256                \\ \hline
	Type-1 Epochs  & 100                 & 30                   & 50                  & 50               & 60                   & -                  \\ \hline
	Type-2 Epochs  & 50                  & 50                   & -                   & -                & 20                   & 10                 \\ \hline
	Gradient Shape & $ 1,024 \times 84 $ & $ 1,024 \times 512 $ & $ 1,436 \times 16 $ & $ 105 \times 12$ & $ 4096 \times 512 $  & -                  \\ \hline
	Feature Shape  & $ 256 \times 400 $  & $ 256 \times 512 $   & -                   & -                & $ 1024 \times 1024 $ & $ 256 \times 150 $ \\ \hline
\end{tabular}
	\end{minipage}
\vspace{-1mm}
\end{figure*}

To show the wide application range of our proposed mechanisms, we run a series of experiments in different settings, including a variety of datasets, models. Experimental results are compared against a number of existing mechanisms to show the superiority of IMGM.

\subsection{Setup}
\textbf{Datasets and tasks.} We select several typical learning tasks from multiple areas where the data are likely to be sensitive. For computer vision, we have two image classification tasks, respectively on MNIST, CIFAR-10 and SVHN. For data mining, we run classification tasks respectively on Cora \cite{Dua:2019} and Adult\cite{Dua:2019}, which is a small-scale dataset on which one predicts whether the income exceeds a threshold. For text mining, we choose a binary classification task on IMDB \cite{maas-EtAl:2011:ACL-HLT2011} dataset. We consider training data as private.

\textbf{Baselines and metrics.} We compare IMGM with other differential privacy mechanisms dealing with high-dimensional data. The baselines include: classic Gaussian mechanism (cGM)~\cite{dwork2014algorithmic}, analytic Gaussian mechanism (aGM) \cite{balle2018improving}, Matrix Variate Gaussian (MVG) \cite{chanyaswad2018mvg}, Moments accountant \cite{abadi2016deep}, Matrix Mechanism (MM) \cite{li2015matrix}. Since cGM, aGM, MM are for vectorized values, we choose to flatten the matrix to apply them. MM is only applicable to small-scale datasets due to its high complexity. For most of the experiments, we use testing accuracy as the utility metric.

\textbf{Type I: Private training gradients.}
We deploy experiments on datasets including MNIST, CIFAR-10, SVHN, Cora and Adult. Here we set the privacy parameter $ \delta $ to $ 10^{-5} $ and choice of the $ \varepsilon $ is in the range of $0.01 \sim 1.0$. We implement the differentially private SGD algorithm, of which the procedures are as follows:

1) Take a random sample from the training set with sampling probability $q$.

2) Compute the gradients on this sample.

3) Clip the gradients by its $l_2$ norm.

4) Average the gradients for a batch of samples, and \textbf{apply perturbation} to the averaged gradients.

5) Update the corresponding model parameters with the perturbed average gradients and go back to 1).

Note that the query here is an average function on gradients. Overall setups and hyperparameters can be found in Tab.~\ref{tab:exp_setups}. We set $ s_2(f) $ as the clip value in IMGM, MVG and MM. Following the convention, we perturb and upload gradients for every lot of training data and the total privacy budget depends on the lot size. For example, we add noise (consume privacy budget) once for training every $1024$ instances on MNIST. We apply the composition theorem from Theorem 3.4 in \cite{kairouz2017composition}. 

Step 4) is where we implement DP mechanisms.  In MM, it solves an optimization problem to find the sum squared error of $WX+A^{+}\|A\|Z$ where $WX$ is the query function. $A$ is the solution to the optimization problem called strategy matrix, and $ A^{+} $ the Moore–Penrose pseudoinverse of $ A $. Since the query here is an average function we set $ W $ as the $ \frac{1}{n}(1,1,\ldots,1)  $ to get the strategy matrix $ A $. In MVG, we choose the \textit{binary precision allocation strategy} \cite{chanyaswad2018mvg} to decide the importance of different directions and SVD to calculate the directional matrix. 

\textbf{Type II: Private training features.}
Four datasets are adopted: MNIST, CIFAR-10, SVHN, and IMDB. We aim to protect the intermediate features of the training data. We modify each model by replacing its activation function with $\tanh(\cdot)$ to normalize the released intermediate-layer feature for untrusted post-processing. For all datasets, stochastic gradient descent is adopted as the optimizer and in each iteration, a batch of $256$ input instances are randomly selected to train the optimizer. Configuration details are given in Table~\ref{tab:exp_setups}.

The released features are perturbed once before training. MM is omitted for not suitable to large-scale datasets. We use the sampling amplification scheme \cite{bassily2014private} for MVG, aGM and IMGM to amplify the privacy budget as $ (\varepsilon/q, \delta/q) $ with sampling rate $ q $.


\begin{figure*}[ht]
	\vspace{-5mm}
	\centering
	\subfigure{
		\begin{minipage}[t]{0.25\linewidth}
			\centering
			\includegraphics[width=1\linewidth ]{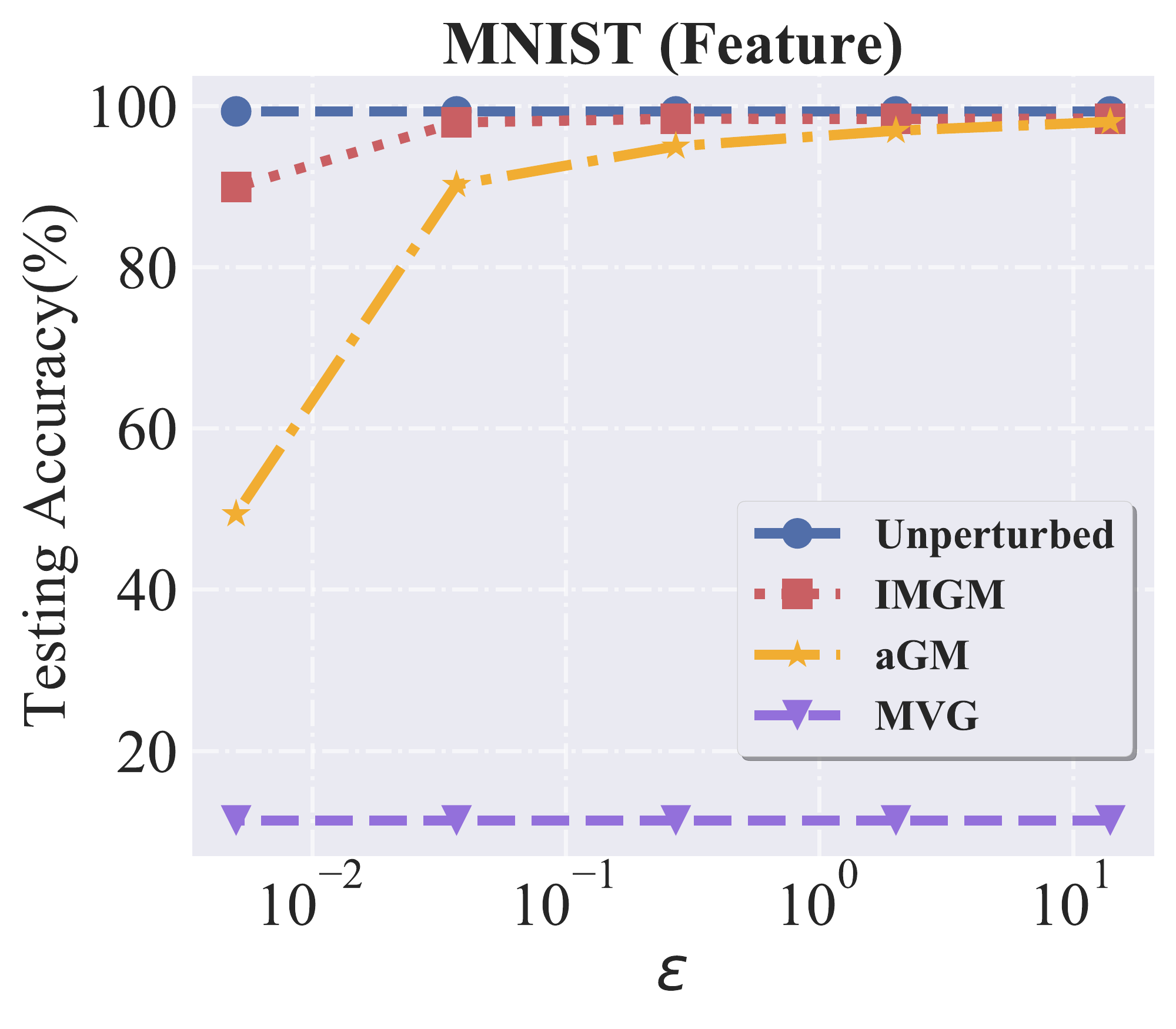}
		\end{minipage}%
	}%
	\subfigure{
		\begin{minipage}[t]{0.25\linewidth}
			\centering
			\includegraphics[width=1\linewidth ]{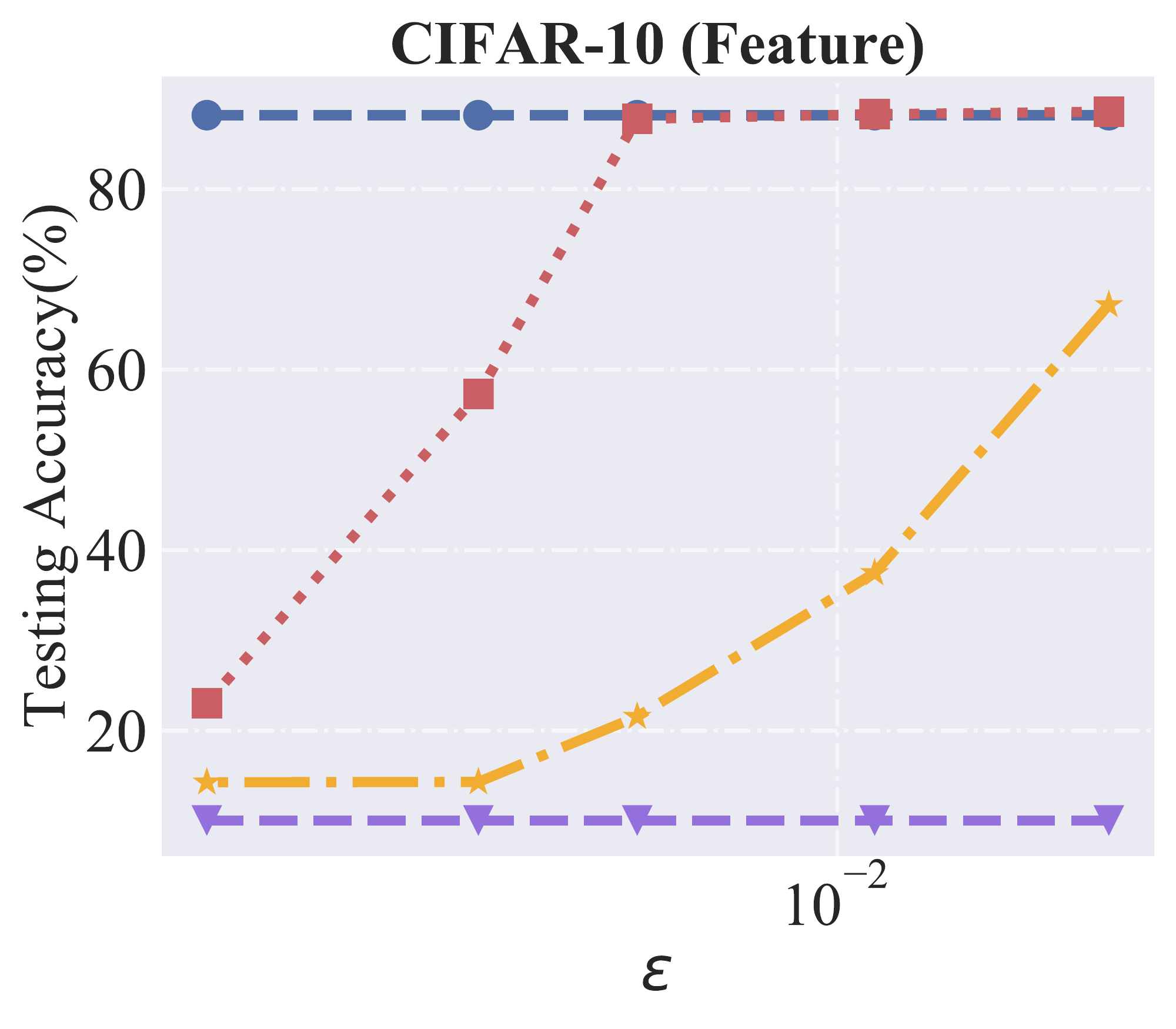}
		\end{minipage}%
	}%
	\subfigure{
		\begin{minipage}[t]{0.25\linewidth}
			\centering
			\includegraphics[width=1\linewidth ]{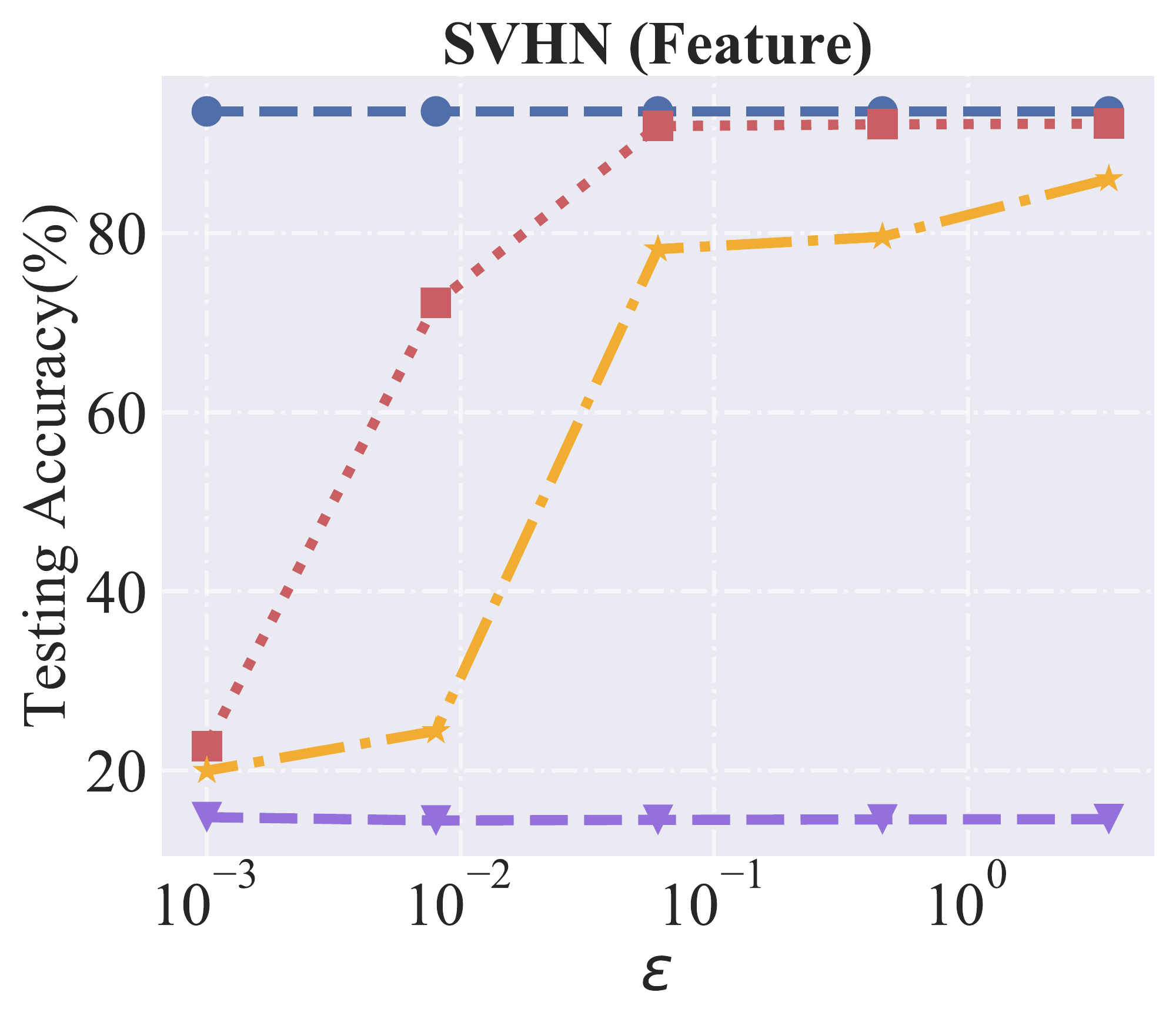}
		\end{minipage}%
	}%
	\subfigure{
		\begin{minipage}[t]{0.25\linewidth}
			\centering
			\includegraphics[width=1\linewidth ]{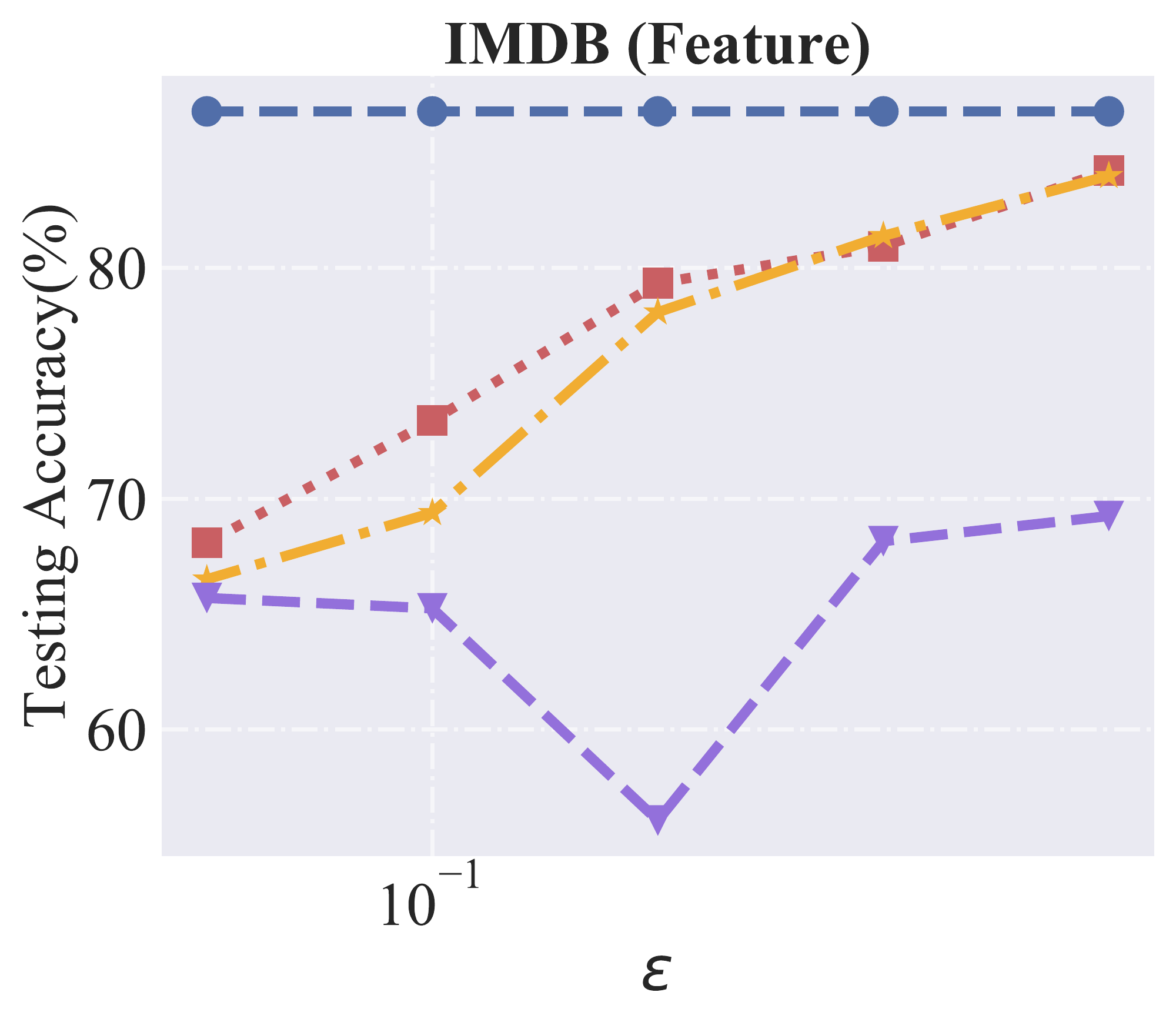}
		\end{minipage}%
	}%
	\vspace{-5mm}
	\caption{Type II private training features results. Legends are shared. IMGM performs best overall, followed by the aGM. }
	\label{fig:train-protection}
	\vspace{-5mm}
\end{figure*}

\subsection{Results}

The results of Type I experiment are in progress.

\textbf{Type II Results.} Fig.~\ref{fig:train-protection} reports accuracies under different $\epsilon$s when fixing $ \delta = 10^{-5}.$  `Unperturbed' represents the case with no privacy guarantee. Overall, the accuracy of Type II is inferior to that of Type I, mostly because noise is inserted at one time, tweaking the original inputs. And the algorithm does not have the opportunity to adapt to the noise as in Type I. Nonetheless, IMGM still outperforms aGM and MVG. In the experiment on IMDB, the performance of aGM is slightly better than IMGM at some large $ \epsilon$s. After all , aGM applies the same necessary and sufficient condition DP but not specific to matrix-valued data. MVG introduces an overwhelming amount of noise which leads to almost random results.

%% file: related.tex
\section{Related work}
Early works on differential privacy were done on small datasets. The standard algorithm for achieving $ (\varepsilon, \delta) $-differential privacy adds random noise from Gaussian distribution \cite{dwork2014algorithmic}. By using numerical evaluations of the Gaussian cumulative density function (CDF) to obtain the optimal variance, analytical Gaussian mechanism \cite{balle2018improving} presents the necessary and sufficient condition of $ (\varepsilon, \delta) $-differential privacy for scalar/vector query. Although it obtains the tightest noise bound ever, it does not provide a solution to the matrix-valued case. 

The Matrix-Variate Gaussian Mechanism \cite{chanyaswad2018mvg} offers a sufficient condition of $ (\varepsilon, \delta) $-differential privacy for matrix-valued queries. It points out the directional noise should be constructed to achieve high utility, yet without a closed-form solution.
The Matrix Mechanism \cite{li2015matrix,mckenna2018optimizing} is a technique for answering
counting queries over vectors while trying to minimize the sum squared error. It solves an optimization problem to find a strategy matrix to adjust the direction of the noise. 

Both IMGM and MVG are the Gaussian mechanisms for matrix-valued queries. However, with a necessary and sufficient condition of DP, we present a better noise direction, {\em i.e.,} i.i.d. Gaussian noise with $\ell_2$-sensitivity-bounded noise variance. Unlike MVG and MM, IMGM merely requires a numerical evaluation of Gaussian CDF, which is far efficient, and achieves high utility, especially in the high dimensional and high privacy regime.

%% file: conclusion.tex
\section{Conclusion}
We propose IMGM, a new differential privacy mechanism for matrix-valued queries. We develop IMGM by transforming the necessary and sufficient condition of $ (\varepsilon,\delta) $-differential privacy in the matrix case, and derive its composition theorem accordingly. We show that the optimal additive Gaussian mechanism for matrices only requires i.i.d. Gaussian noise, which is different from previous works proposing directional noise. Hence such noise is efficient to obtain in practice. 

%% file: appendix2.tex
\section{Necessary and Sufficient Condition for DP}

\subsection{Proof of Lemma \ref{privacy loss}}\label{appendix: lemma 1}
\begin{proof}
The privacy loss random variable $L_{K, X, X^{\prime}}=\ell_{K, X, X^{\prime}}(Y)$ is the transformation of the output random variable $Y=K(X)$ by the function $\ell_{K, X, X^{\prime}}$. For the particular case of a Matrix Gaussian mechanism $K(X)=f(X)+Z$ with ${Z} \sim  \mathcal{MN}_{m,n}(\mathbf{0},\Sigma_{1},\Sigma_{2})$ where $ U_{k}U_{k}^{\top} = \Sigma_{k}, ~ k=1,2$, and $\Delta = f(X) - f(X^{\prime})$, we have

	\begin{equation}
	\begin{split}
		&\ell_{K, X, X^{\prime}}(Y)=\log \left(\frac{p_{K(X)}(Y)}{p_{K\left(X^{\prime}\right)}(Y)}\right)\\
		=&\log \frac{\exp( -\frac{1}{2} \| U_1^{-1} {Z} U_2^{-\top}\|_F^2 )}{\exp( -\frac{1}{2} \| U_1^{-1} ({Z}+\Delta) U_2^{-\top}\|_F^2 )} \\
		=& \frac{1}{2}\| U_1^{-1} ({Z}+\Delta) U_2^{-\top}\|_F^2 -\frac{1}{2}\| U_1^{-1} {Z} U_2^{-\top}\|_F^2\\
		= & \frac{1}{2} \| \Delta^{\prime} \|_F^2 + vec({Z}^{\prime})^{\top} vec(\Delta^{\prime})
	\end{split}
	\end{equation}
	where $ {\Delta}^{\prime} = U_1^{-1} {\Delta} U_2^{-\top},$ $ {{Z}}^{\prime} = U_1^{-1}{Z}U_2^{-\top}.$ Since $ {Z}\sim \mathcal{MN}_{m,n}(\mathbf{0},\Sigma_{1},\Sigma_{2})$, we have $ {{Z}}^{\prime} \sim \mathcal{MN}_{m,n}(\mathbf{0},\mathbf{E}_{1},\mathbf{E}_{2}),$ and $ vec({Z}^{\prime})^{\top} vec(\Delta^{\prime}) \sim \mathcal{N}(0,\| \Delta^{\prime} \|_F^2) $. Letting $ \eta = \frac{1}{2} \| \Delta^{\prime} \|_F^2  $, we could get that $ L_{K, X, X^{\prime}} $ follows a Gaussian distribution $\mathcal{N}(\eta, 2 \eta)$. Note that $\eta$ actually depends on $X, X^{\prime}$.
\end{proof}

\subsection{Proof of Lemma \ref{lemma: IMGM}}\label{appendix: lemma 3}
\begin{proof}
	Suppose $K(X)=f(X)+Z$ is a Gaussian output perturbation mechanism with $Z \sim  \mathcal{MN}_{m,n}(\mathbf{0},\Sigma_{1},\Sigma_{2})$. For any datasets $X \simeq X^{\prime}$, let $\Delta=f(X)-f\left(X^{\prime}\right).$ Then the following holds for any $\varepsilon \geq 0$ :
	\begin{equation}\label{eq: privacy loss Phi part_1}
	\begin{split}
	\Pr \left[ L_{K, X, X^{\prime}} \geq \varepsilon\right] & = \Pr \left[ \mathcal{N}(\eta, 2\eta) \geq \varepsilon\right] \\
	& = \Pr \left[ \mathcal{N}(0, 1) \geq \frac{\varepsilon - \eta}{\sqrt{2\eta}}  \right]\\
	& = \Pr \left[ \mathcal{N}(0, 1) \leq \frac{\eta - \varepsilon}{\sqrt{2\eta}}  \right]\\
	& = \Phi\left(\frac{ \|\Delta^{\prime}\|_F}{2}-\frac{\varepsilon }{\|\Delta^{\prime}\|_F}\right).
	\end{split}
\end{equation}
With a symmetric argument, we have
	\begin{equation}\label{eq: privacy loss Phi part_2}
	\Pr\left[L_{K, X^{\prime}, X} \leq-\varepsilon\right]=\Phi\left(-\frac{\|\Delta^{\prime}\|_F}{2 }-\frac{\varepsilon }{\|\Delta^{\prime}\|_F  }\right),
	\end{equation}
	where $ \Delta^{\prime} = U_1^{-1} \Delta U_2^{-\top}$ and $ U_k U_k^{\top} = \Sigma_{k},~k=1,2 .$

	By substituting Eq.~\eqref{eq: privacy loss Phi part_1} and Eq.~\eqref{eq: privacy loss Phi part_2} into Eq. \ref{eq: lemma privacy loss inequality}, we can get $g(\|\Delta^{\prime}\|_F)$ as the same $g(\cdot)$ in Eq.~\eqref{eq: Phi}. Note that the left side of Eq.~\eqref{eq: Phi} is a monotonically increasing function of $ x $, and the range of left side of Eq.~\eqref{eq: Phi} is $ (0,1) $. So there must exist a root $ B $ for the equality of Eq.~\eqref{eq: Phi} to hold. Therefore, for $X \simeq X^{\prime}$, the inequality~\eqref{eq: lemma privacy loss inequality} is the same as $\|\Delta^{\prime}\|_F \le B.$
\end{proof}

\subsection{Proof of Thm.~\ref{thm: IMGM}}\label{appendix: thm 1}

We first provide the following lemma and its proof.
\begin{lemma}\label{lemma: norm singular value inequality}
	For the matrix $ A\in \mathbb{R}^{m\times m}, B \in \mathbb{R}^{m\times n}, C \in \mathbb{R}^{n \times n}$, $ \{\sigma_i(A)\}_{i=1}^{m} $ is a non-increasing sequence which represents the Singular Value of matrix $ A $, so as $ B,~C $. Then we have
	\begin{equation}
		\|ABC\|_F^2 \le \sum_{i=1}^{r} \sigma_i^2(A)\sigma_i^2(B) \sigma_i^2(C)
	\end{equation}
	where $ \|\cdot\|_F $ is the Frobenius norm and $ r = \min\{m,n\} $.
\end{lemma}
\begin{proof}
	We need to apply Singular Value Decomposition (SVD) in first step. The matrix $ A,B,C $ can be expressed in the form as 
	\begin{equation}
	A = U_A \Sigma_A V_A^{\top},~B = U_B \Sigma_B V_B^{\top},~C = U_C \Sigma_C V_C^{\top},
	\end{equation}
	where $ U_A, V_A, U_B, V_B, U_C,V_C $ are orthogonal matrices and 
	 \begin{equation}
	 \begin{split}
	 	 \Sigma_A=diag(\sigma_1(A),\ldots,\sigma_m(A)),\\ \Sigma_B=diag(\sigma_1(B),\ldots,\sigma_r(B)),\\ 
	 	 \Sigma_C=diag(\sigma_1(C),\ldots,\sigma_n(C)),
	 \end{split}
	 \end{equation}
	are diagonal matrices.
	Since the Frobenius norm will not change with the orthogonal transform, we could get 
	\begin{equation}
	\begin{split}
		\|ABC\|_F^2 &= \|U_A \Sigma_A V_A^{\top}U_B \Sigma_B V_B^{\top}U_C \Sigma_C V_C^{\top}\|_F^2 \\
		& = \|\Sigma_A U \Sigma_B V \Sigma_C \|_F^2
	\end{split}
	\end{equation}
	where $ U = (u_{ij})_{m\times m}= V_A^{\top} U_B , V = (v_{ij})_{n \times n} = V_B^{\top}U_C $ are orthogonal matrices. Hence, 
	\begin{equation}
	\begin{split}
		\|ABC\|_F^2 = &\sum_{i=1}^{m}\sum_{j=1}^{n} \left[\sum_{k=1}^{r} \sigma_i(A) \sigma_k(B) \sigma_j(C) u_{ik}v_{kj}\right]^2\\
		= & \sum_{i=1}^{m}\sum_{j=1}^{n} \sigma_i^2(A) \sigma_j^2(C)\left[\sum_{k=1}^{r} \sigma_k(B) u_{ik}v_{kj}\right]^2.
	\end{split}
	\end{equation}
	We can also observe that 
	\begin{equation}
	\|U\Sigma_B V\|_F^2 = \sum_{i=1}^{m}\sum_{j=1}^{n} \left[\sum_{k=1}^{r} \sigma_k(B) u_{ik}v_{kj}\right]^2.
	\end{equation}
	As U, V are orthogonal matrices, we have
	\begin{equation}
	\|U\Sigma_B V\|_F^2 = \|\Sigma_B\|_F^2 = \sum_{k=1}^{r}\sigma_k^2(B).
	\end{equation}
	Let $ b_{ij} = \sum_{k=1}^{r} \sigma_k(B) u_{ik}v_{kj} $, 
	\begin{equation}
	\sum_{i=1}^{m}\sum_{j=1}^{n}  b_{ij}^2 = \|U\Sigma_B V\|_F^2  =\sum_{k=1}^{r}\sigma_k^2(B).
	\end{equation}
And 
\begin{equation*}
\|ABC\|_F^2 = \sum_{i=1}^{m}\sum_{j=1}^{n} \sigma_i^2(A) \sigma_j^2(C)b_{ij}^2.
\end{equation*}
Then we will prove
\[\sum_{i=1}^{m}\sum_{j=1}^{n} \sigma_i^2(A) \sigma_j^2(C)b_{ij}^2 \le  \sum_{i=1}^{r} \sigma_i^2(A)\sigma_i^2(B) \sigma_i^2(C) . \] 
First, we know 
\[
\sigma_1(A) \geq \cdots \geq \sigma_m(A) \geq 0,~
\sigma_1(C) \geq \cdots \geq \sigma_n(C) \geq 0.
\]
Then there exist non-negative numbers $ \xi_t, \eta_s (1\le t\le m,~1\le s \le n)$ such that
\begin{equation}
\sigma_i^2(A)=\sum_{t = i}^{m} \xi_t,~~\sigma_j^2(C)=\sum_{s = j}^{n} \eta_s.
\end{equation}
Hence, we use the symbol $ \delta_{ij} $ denote as 
\begin{equation}
\delta_{ij} = \left\{
\begin{aligned}
&\sigma_i(B),~~i = j \\
&0,~~~~~~~~i \neq j
\end{aligned}
\right.
\end{equation}
we have 
\begin{equation}\label{sum exchange}
\begin{split}
&\sum_{i=1}^{r} \sigma_i^2(A)\sigma_i^2(B) \sigma_i^2(C) - \sum_{i=1}^{m}\sum_{j=1}^{n} \sigma_i^2(A) \sigma_j^2(C)b_{ij}^2 \\
= & \sum_{i=1}^{m}\sum_{j=1}^{n}  (\delta_{ij}^2 - b_{ij}^2) \sigma_i^2(A) \sigma_j^2(C)  \\
= & \sum_{i=1}^{m}\sum_{j=1}^{n} (\delta_{ij}^2 - b_{ij}^2) \sum_{t = i}^{m} \xi_t \sum_{s = j}^{n} \eta_s \\
= & \sum_{t = 1}^{m}  \sum_{s = 1}^{n} \xi_t \eta_s \sum_{i=1}^{t}\sum_{j=1}^{s} (\delta_{ij}^2 - b_{ij}^2).
\end{split}
\end{equation}
If $ t \le s $, then the inner sum on  the right hand side of (\ref{sum exchange}) is non-negative since
\begin{equation}\label{final equation holds}
\sum_{i=1}^{t}\sum_{j=1}^{s} (\delta_{ij}^2 - b_{ij}^2) \geq \sum_{i=1}^{t}\sum_{j=1}^{n} (\delta_{ij}^2 - b_{ij}^2) = 0.
\end{equation}
The final equation holds due to 
\begin{equation}
\sum_{i=1}^{t}\sum_{j=1}^{n} \delta_{ij}^2 = \sum_{k=1}^{t} \sigma_k^2(B).
\end{equation}
And we set a matrix as
\begin{equation}
E(t) = \left(
\begin{array}{cc}
E_{t, t} & 0_{t, n-t} \\
0_{n-t, t} & 0_{n-t,n-t}\\
\end{array}
\right)
\end{equation}
where $ E_{t, t} $ is an identity matrix with order $ t $ and other elements are all $ 0 $. Hence, we have 
\begin{equation}
\begin{split}
\sum_{i=1}^{t}\sum_{j=1}^{n} b_{ij}^2 & = \|U\Sigma_B V E(t)\|_F^2 \\
& = \|U\Sigma_B E(t) V \|_F^2 \\&= \|\Sigma_B E(t) \|_F^2 
= \sum_{k=1}^{t} \sigma_k^2(B).
\end{split}
\end{equation}
Therefore, Eq. \ref{final equation holds} holds, which completes the proof.
\end{proof}

With the lemma~\ref{lemma: norm singular value inequality}, we could give the proof of Thm.~\ref{thm: IMGM}.
\begin{proof}
	From Lemma~\ref{lemma: IMGM}, we know that the inequality~\eqref{eq: lemma privacy loss inequality} is equivalent to $ \|\Delta^{\prime}\|_F \le B $ for every $X \simeq X^{\prime}$. And $ \Delta^{\prime} = U_1^{-1} \Delta U_2^{-\top} $. Hence, we could apply Lemma \ref{lemma: norm singular value inequality} to $ \|\Delta^{\prime}\|_F  $ and obtain
	\begin{equation}
	\|\Delta^{\prime}\|_F^2 \le \sum_{i=1}^{r}  \sigma_i^2(U_1^{-1})\sigma_i^2(\Delta) \sigma_i^2(U_2^{-1}).
	\end{equation}
	We design $ U_1 = W_{U_1} S_{U_1} $ where $ W_{U_1} $ is an orthogonal matrix and $ S_{U_1} = diag(\sigma_{1}(U_1), \ldots,\sigma_{m}(U_1)) $ is a diagonal matrix. We set the order from $ m $ to $ 1 $ so to keep $ \frac{1}{\sigma_{i}(U_1)} $ a non-increasing sequence, wihch follows the definition of SVD.  Therefore, $ S_{U_1}^{-1} = diag(\frac{1}{\sigma_{m}(U_1)}, \ldots, \frac{1}{\sigma_{1}(U_1)}) $, and we do the same for $ U_2 $ and $ \sigma_i(U_2) = \sigma_i({U_2^{\top}}) $. Hence, 
	\begin{equation}\label{eq: the upper bound of Delta'}
	\|\Delta^{\prime}\|_F^2 \le \sum_{i=1}^{r} \frac{\sigma_i^2(\Delta)}{\sigma_{m-i+1}^2(U_1)\sigma_{n-i+1}^2(U_2)}.
	\end{equation}
	It is clear that if $$\sum_{i=1}^{r} \frac{\sigma_i^2(\Delta)}{\sigma_{m-i+1}^2(U_1)\sigma_{n-i+1}^2(U_2)} \le B^{2},$$ we must have $\|\Delta^{\prime}\|_F^2 \le B^{2}$. Proof completes.
\end{proof}

\subsection{Proof of Thm.~\ref{thm: IMGM_new}}\label{appendix: proof of thm IMGM new}
\begin{proof}
	
We will prove the theorem by two parts. 

First, $(\varepsilon, \delta)$-DP $ \Longrightarrow  $ Thm.~\ref{thm: IMGM_new}.
By Lemma~\ref{lemma: privacy loss iff DP}, we know the necessary and sufficient condition for $(\varepsilon, \delta)$-DP is such that, for any adjacent input pair $ X $ and $ X^{\prime}, $
\begin{equation}\label{eq: appendix privacy loss inequality}
\Pr \left[L_{K, X, X^{\prime}} \geq \varepsilon\right]-e^{\varepsilon} \Pr \left[L_{K, X^{\prime}, X} \leq-\varepsilon\right] \leq \delta.
\end{equation}
By Lemma~\ref{lemma: IMGM}, it is equivalent to have Eq.~\ref{eq:maxdp} hold true. Since we have Eq.~\ref{eq: the upper bound of Delta'} and the upper bound can be achieved by setting $U_1 =  W_{U_1}S_{U_1} $ and $U_2 =  W_{U_2}S_{U_2}$, we further have 
\begin{multline}\label{eq:ineqsigma}
\sum_{i=1}^{r} \frac{\sigma_i^2(\Delta)}{\sigma_{m-i+1}^2(U_1)\sigma_{n-i+1}^2(U_2)} \le \sum_{i=1}^{r} \frac{\sigma_i^2(\Delta)}{\sigma_{m}^2(U_1)\sigma_{n}^2(U_2)} \\
\le \frac{s_{2}^{2}(f)}{\sigma_{m}^2(U_1)\sigma_{n}^2(U_2)}.
\end{multline}
And the equality can be achieved if we choose the $ \Delta $ subject to $ \{\sigma_{i}^2(\Delta)\}_{i=1}^{r} = (s^2_2(f),0,0,\ldots,0) .$ Hence, if the $(\varepsilon, \delta)$-DP condition holds, for any adjacent input pairs, we have 
\begin{equation}
\frac{s^2_2(f)}{\sigma_{m}^2(U_1) \sigma_{n}^2(U_2)} \le B^2,
\end{equation}
which we could rewrite as:
\begin{equation}
\sigma_{m}^2(U_1) \sigma_{n}^2(U_2) \geq \frac{s^2_2(f)}{B^2}.
\end{equation}

Second, $(\varepsilon, \delta)$-DP $ \Longleftarrow  $ Thm.~\ref{thm: IMGM_new}. We will prove that Thm.~\ref{thm: IMGM_new} will lead to the condition in Lemma~\ref{lemma: privacy loss iff DP}. If Thm.~\ref{thm: IMGM_new} holds, we have for any adjacent input pair $ X $ and $ X^{\prime} $,
\begin{equation}\label{eq: thm3 -> thm 2}
\begin{split}
& \sum_{i=1}^{r}\frac{\sigma_i^2(\Delta)}{\sigma_{m-i+1}^2(U_1)\sigma_{n-i+1}^2(U_2)}  \\
\le  & \sum_{i=1}^{r}\frac{\sigma_i^2(\Delta)}{s^2_2(f)/B^2} \\ 
= & \frac{B^2}{s^2_2(f)}\sum_{i=1}^{r} \sigma_i^2(\Delta) \\
\le & B^2.
\end{split}
\end{equation}
The first inequality holds since
\begin{equation}
\begin{split}
\sigma_{m-i+1}^2(U_1) \sigma_{n-i+1}^2(U_2) \geq  &
\sigma_{m}^2(U_1) \sigma_{n}^2(U_2)  \\ 
& \geq \frac{s^2_2(f)}{B^2}, ~ i \in [r],
\end{split}
\end{equation}
with the condition that the singular values $\{ \sigma_i(U_1) \}_{i=1}^{m}$ and $\{ \sigma_i(U_2) \}_{i=1}^{n}$ is the non-increasing. And the last inequality holds due to $ \sum_{i=1}^{r} \sigma_i^2(\Delta) = \|\Delta\|_F^2 \le s_2^2(f) $. By Thm.~\ref{thm: IMGM}, Eq.~\eqref{eq: thm3 -> thm 2} leads to 
\begin{equation}
\| U_{1}^{-1}\Delta U_2^{-\top} \|_F \le B.
\end{equation}
With Lemma~\ref{lemma: IMGM} and $ \Delta^{\prime} = U_{1}^{-1}\Delta U_2^{-\top}  $, we have
\begin{equation}
g(\|\Delta^{\prime}\|_F) \le \delta,
\end{equation} 
for any adjacent input pair $ X $ and $ X^{\prime} $. It leads to $(\varepsilon, \delta)$-DP by using Lemma~\ref{lemma: privacy loss iff DP}. Proof completes.
\end{proof}

\subsection{Algorithm to Caculate $ B $}\label{appendix: alg to  Caculate $ B $}

We present Alg.~\ref{alg: privacy bound} to caculate $ B $ in Lemma~\ref{lemma: IMGM}, in which the main procedure is to use numerical calculation to solve the transcendental equation that $ g(x) = \delta $. Thus a numerical solution $ \hat{B} $ is given for $B$.
\begin{algorithm}
	\renewcommand{\algorithmicrequire}{\textbf{Input:}}
	\renewcommand{\algorithmicensure}{\textbf{Output:}}
	\caption{Caculate Privacy Bound}
	\label{alg: privacy bound}
	\begin{algorithmic}[1] 
		\REQUIRE (a) privacy parameters $\varepsilon,\delta$.
		\ENSURE privacy bound $ \hat{B} $ and the variance $ \sigma $
		\STATE {Let $\delta_{0}=\Phi(0)-e^{\varepsilon} \Phi(-\sqrt{2 \varepsilon})$}
		\IF {$\delta \geq \delta_{0}$} 	 
		\STATE { Define $B_{\varepsilon}^{+}(v)=\Phi(\sqrt{\varepsilon v})-e^{\varepsilon} \Phi(-\sqrt{\varepsilon(v+2)})$ }
		\STATE {Compute $v^{*}=\sup \left\{v \in \mathbb{R}_{\geq 0}: B_{\varepsilon}^{+}(v) \leq \delta\right\}$}
		\STATE {Let $\alpha=\sqrt{1+v^{*} / 2}-\sqrt{v^{*} / 2}$}
		\ELSE	
		\STATE { Define $B_{\varepsilon}^{-}(u)=\Phi(-\sqrt{\varepsilon u/T})-e^{\varepsilon} \Phi(-\sqrt{\varepsilon(u+2)})$ }
		\STATE {Compute $u^{*}=\inf \left\{u \in \mathbb{R}_{\geq 0}: B_{\varepsilon}^{-}(u) \leq \delta\right\}$}
		\STATE {Let $\alpha=\sqrt{1+u^{*} / 2}+\sqrt{u^{*} / 2}$}
		\ENDIF
		\STATE {Let $ \hat{B} = \sqrt{2\varepsilon}/\alpha $}
		\RETURN {$\hat{B} $}
	\end{algorithmic} 
\end{algorithm}

\subsection{Proof of Composition}\label{appendix: proof of deep learning with IMGM}

Here we present the proof of Thm.~\ref{thm: IMGM_Deep learning}.
\begin{proof}
	 We assume all the mechaisms $ {K}_{1}, \ldots, {K}_{T} $ are independent and identical, with the same $ l_2 $-sensitivity for each $ t \in [T] $. Therefore, every privacy loss follows $ \mathcal{N}(\eta , 2\eta ) $ by Lemma~\ref{privacy loss}. For any $X \simeq X^{\prime}$, we compose over their privacy loss variables:
	\begin{equation}
	L_{K,X,X^{\prime}} = \sum_{t=1}^{T}  L_{K_{t}, X, X^{\prime}}
	\end{equation}
	Due to the additive property of the Gaussian variable, the distribution of $ L_{K,X,X^{\prime}}  $ is $ \mathcal{N}(T\eta , 2T\eta ).$ Following a similar proof procedure in Lemma~\ref{lemma: IMGM}, we could get 
	\begin{equation}
	\begin{split}
	\Pr \left[ L_{K, X, X^{\prime}} \geq \varepsilon\right] & = \Pr \left[ \mathcal{N}(T\eta, 2T\eta) \geq \varepsilon\right] \\
	& = \Pr \left[ \mathcal{N}(0, 1) \leq \frac{T\eta  - \varepsilon}{\sqrt{2T\eta }}  \right]\\
	& = \Phi\left(\frac{ \sqrt{T}\|\Delta^{\prime}\|_F}{2}-\frac{\varepsilon }{\sqrt{T}\|\Delta^{\prime}\|_F}\right).
	\end{split}
	\end{equation}
	A symmetric argument is
	\begin{equation}
	\Pr\left[ L_{K, X, X^{\prime}} \leq -\varepsilon\right] = \Phi\left(-\frac{ \sqrt{T}\|\Delta^{\prime}\|_F}{2}-\frac{\varepsilon }{\sqrt{T}\|\Delta^{\prime}\|_F}\right).
	\end{equation}
	Therefore, by Lemma~\ref{lemma: privacy loss iff DP} the composed mechanism $ K $ satisfies Eq.~\ref{eq: lemma privacy loss inequality} for a particular pair of $X \simeq X^{\prime}$ if and only if
	\begin{equation}\label{eq: Phi with T}
	\resizebox{.91\linewidth}{!}{$
		\displaystyle
		\Phi\left(\frac{\sqrt{T}\|\Delta^{\prime}\|_F}{2}-\frac{\varepsilon }{\sqrt{T}\|\Delta^{\prime}\|_F}\right)-e^{\varepsilon} \Phi\left(-\frac{\sqrt{T}\|\Delta^{\prime}\|_F}{2 }-\frac{\varepsilon }{\sqrt{T}\|\Delta^{\prime}\|_F  }\right) \leq \delta.
		$}
	\end{equation}
	Following the same reason in the proof of Lemma.~\ref{lemma: IMGM}, we have an equivalent condition of Eq.~\eqref{eq: Phi with T} such that 
	\begin{equation}
	\sqrt{T}\|\Delta^{\prime}\|_F \le B,
	\end{equation}
	where $ B $ holds the same value as in Thm.~\ref{thm: IMGM}. The rest argument is omitted as it is similar to the proof without composition. Hence the necessary and sufficient condition for $(\varepsilon, \delta)$-DP on the composed mechanism $K$ is that:
	\begin{equation}
	\sigma_{m-i+1}(U_1) \sigma_{n-i+1}(U_2) \geq \frac{ s_2(f) \sqrt{T}}{B},~\forall i \in [r] .
	\end{equation}
\end{proof}

\section{Connection with R\'{e}nyi Differential Privacy}
We prove that IMGM is a general mechanism which also satisfies $ (\alpha, \varepsilon) $-R\'{e}nyi Differential Privacy (RDP) \cite{mironov2017renyi}, and can be directly applied composition such as Moments Account \cite{abadi2016deep} as a primitive mechanism.

For any pair of adjacent datasets ${X}$ and ${X}^{\prime}$, the R\'{e}nyi divergence between query results on the two datasets is
\begin{equation}
\begin{split}
&D_{\alpha}  ( \Pr( f({X} ) + {Z} \in  {\mathcal{O}}  )  \|\Pr(f({X}^{\prime}) + {Z} \in  {\mathcal{O}}   ) ) \\
=&\frac{1}{\alpha - 1}  \mathbb{E}_{\Pr(f({X}^{\prime}))} \left(\frac{\Pr( f({X} ) + {Z} \in  {\mathcal{O}}  ) }
{\Pr(f({X}^{\prime}) + {Z} \in  {\mathcal{O}}   )} \right)^{\alpha}\\
= &\frac{1}{\alpha - 1}  \mathbb{E}_{\Pr(f({X}^{\prime}))}
\left[  \ell_{K, X, X^{\prime}}(Y) \right]^{\alpha}\\
=  &\frac{(2\alpha+1) \eta}{\alpha - 1}  e^{\eta(\alpha^2+\alpha)} \\
\le & \frac{(2\alpha+1) B}{2(\alpha - 1)}  e^{B(\alpha^2+\alpha)/2}
\end{split}
\end{equation}
where $ B $ is defined in Lemma~\ref{lemma: IMGM}.
Therefore, IMGM satisfies $ (\alpha, \varepsilon^{\prime}) $-RDP, where $ \varepsilon^{\prime} =  \frac{(2\alpha+1) B}{2(\alpha - 1)}  e^{B(\alpha^2+\alpha)/2} $.